\documentclass[conference,compsoc]{IEEEtran}
\ifCLASSOPTIONcompsoc
  \usepackage[nocompress]{cite}
\else
  \usepackage{cite}
\fi
\ifCLASSINFOpdf
   \usepackage[pdftex]{graphicx}
   \DeclareGraphicsExtensions{.pdf,.jpeg,.png}
\else
   \usepackage[dvips]{graphicx}
   \graphicspath{{../eps/}}
   \DeclareGraphicsExtensions{.eps}
\fi
\usepackage{amsmath}
\usepackage{algorithm}

\usepackage{algpseudocode}
\begin{document}
\title{Discovering Technology Gaps using the IntSight Knowledge Navigator}

% \author{\IEEEauthorblockN{Subhasis Dasgupta\\ and Amarnath Gupta}
% \IEEEauthorblockA{University of California\\
% San Diego, CA 92093\\
% Email: sudasgupta, a1gupta@ucsd.edu}
% \and
% \IEEEauthorblockN{Aurpon Gupta}
% \IEEEauthorblockA{Integrative Insights\\
% San Diego, CA 92122\\
% Email: aurpongupta@gmail.com}
% \and
% \IEEEauthorblockN{Snehasis Sinha}
% \IEEEauthorblockA{Walmart\\
% Bangalore, India\\
% Email: snehasis@gmail.com}
% \and
% \IEEEauthorblockN{James Kirk\\ and Montgomery Scott}
% \IEEEauthorblockA{Starfleet Academy\\
% San Francisco, California 96678-2391\\
% Telephone: (800) 555--1212\\
% Fax: (888) 555--1212}
%}


% compsoc conferences compared to traditional conferences), use this
\author{\IEEEauthorblockN{Aurpon Gupta\IEEEauthorrefmark{1},
Subhasis Dasgupta\IEEEauthorrefmark{1}\IEEEauthorrefmark{2},
Snehasis Sinha\IEEEauthorrefmark{3} and 
Amarnath Gupta\IEEEauthorrefmark{1}\IEEEauthorrefmark{2} }
\IEEEauthorblockA{\IEEEauthorrefmark{1}Integrative Insights, San Diego, California 92122 \\
Email: aurpongupta@gmail.com}
\IEEEauthorblockA{\IEEEauthorrefmark{2}University of California San Diego, La Jolla, CA 92093, USA\\
Email: sudasgupta,a1gupta@ucsd.edu}
\IEEEauthorblockA{\IEEEauthorrefmark{3}Walmart India, Bangalore, India 560103
}
}


\maketitle

\begin{abstract}
Knowledge analysis is an important application of knowledge graphs. In this paper, we present a complex knowledge analysis problem that discovers the gaps in the technology areas of interest to an organization. Our knowledge graph is developed on a heterogeneous data management platform. The analysis combines semantic search, graph analytics, and polystore query optimization.
\end{abstract}


% \fi
\IEEEpeerreviewmaketitle

\section{Introduction}
\label{sec:intro}
\begin{figure}[h]
    \centering
    \includegraphics[width=0.45\textwidth]{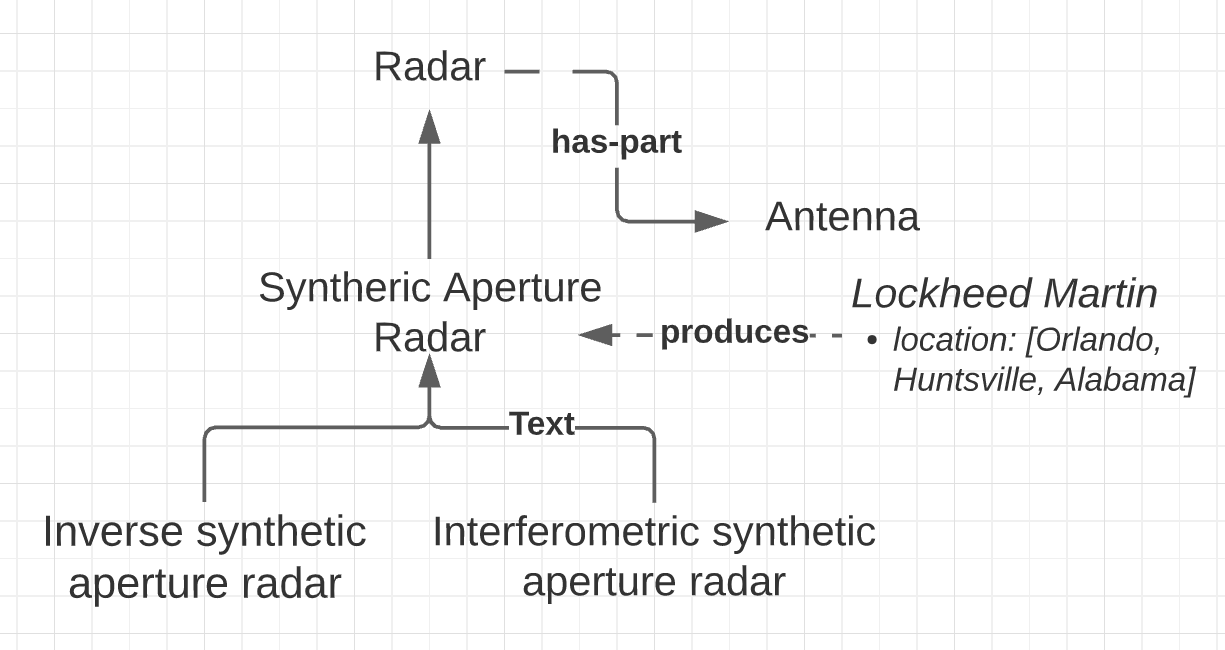}
    %\vspace{-1.5in}
    \caption{A toy knowledge graph}
    \label{fig:smallKG}
\end{figure}

Creation, exploration and management of domain-specific knowledge has become an important research issue both in academia and in the industry \cite{zhou2019agent,leijie2018constructing,wang2017product,zhang2020new}.  The goal of this paper is to present a real-life \textit{knowledge analysis problem} and to introduce IntSight Knowledge Navigator (IKN), a tool designed to address this class of problems. Consistent with today's methodology, the ``knowledge'' in our setting is modeled as a graph whose nodes represent entity classes, instances contain entity properties, and the edges represent class-level, instance-level, and class-membership relationships. We use the term ``entity'' to represent items of interest in an application domain. An item can be a concrete object like a commercially available product (e.g., a GPU), or a more conceptual entity like a technology domain (e.g., microelectronics). Figure \ref{fig:smallKG} shows an example knowledge graph in our domain.

\medskip
\noindent \textbf{The User-Level Problem.} Consider a customer that performs technology research and produces technology products in several different areas. It works with a large number of research and industrial partners who also produce research and/or products in the same space. The company is trying to create a future growth plan in its current areas of expertise as well as some related areas where it is trying to expand. As part of its competitive landscape analysis, it would like to perform a ``technology gap analysis'', which is a discovery process to identify technology areas in which its competitors have made more progress compared to the company itself and its ecosystem of partners. The outcome of the discovery process is to understand the gap areas, the players involved, and the nature of their advancement. 

% \medskip

% \noindent \textbf{Challenges.} 

In this paper, we present a methodology using a combination of semantic search, graph analysis, and polystore optimization. 

\section{The Knowledge Graph Data Architecture}
\label{sec:arch}
The knowledge graph underlying the IntSight Knowledge Network consists of the ontology and the data graph portions, as well as a mapping structure that bridges the two. The data is stored in a commercial version of the polystore \cite{gupta:awesome:2016} built on top of a relational DBMS (PostgreSQL), a graph DBMS (Neo4J) and, text indexes (Apache Solr) to store different portions of the knowledge graph.

% \begin{figure}
%     \centering
%     %\includegraphics{}
%     \vspace{2in}
%     \caption{The ontology graph and its path structure}
%     \label{fig:ontoPath}
% \end{figure}

% \medskip

\noindent \textbf{Ontology.} The ontology, modeled in OWL-DL \cite{motik2004query}, is transformed into a property graph that preserves all ontology properties \cite{gong2018neo4j}. The subproperty axioms in OWL-DL are maintained in a separate tree in the same property graph. Currently, the model does not have any chain rules. Each transitive ontological relationship like \texttt{subclassOf} and \texttt{componentOf} is a directed acyclic graph. Every node maintains a list of the root-to-node tree-path in Apache Solr to avoid explicit graph traversal in the graph database for transitive closure and reachability queries. For nodes with multiple root-to-node paths, we maintain additional paths from the parent of the nearest join-node (of a DAG) to the current node. Further, a term-to-path index is maintained to find relevant paths for a term. 
%Figure \ref{fig:ontoPath} shows the structure for a small ontology. 
\begin{table}[ht]
    \centering
    \begin{tabular}{|c|c|p{1.3in}|} \hline
         \textbf{Data Source}&  \textbf{Data Model} &\multicolumn{1}{c|}{\textbf{Polystore Placement}} \\ \hline
         Patents &Relational  &PostgreSQL, Text in Solr \\ \hline
         News articles &Structured Text  &Solr, Entity Network in Neo4J \\ \hline
         Federal Spending&Relational  &PostgreSQL \\ \hline
         Company Networks&Graph  &Neo4J \\ \hline         
         \end{tabular}
         \vspace{1em}
    \caption{Data from a source are processed and placed into different stores.}
    \label{tab:sources}
    \vspace{-1.3em}
\end{table}

\begin{figure*}
    \centering
  \includegraphics[width=0.9\textwidth]{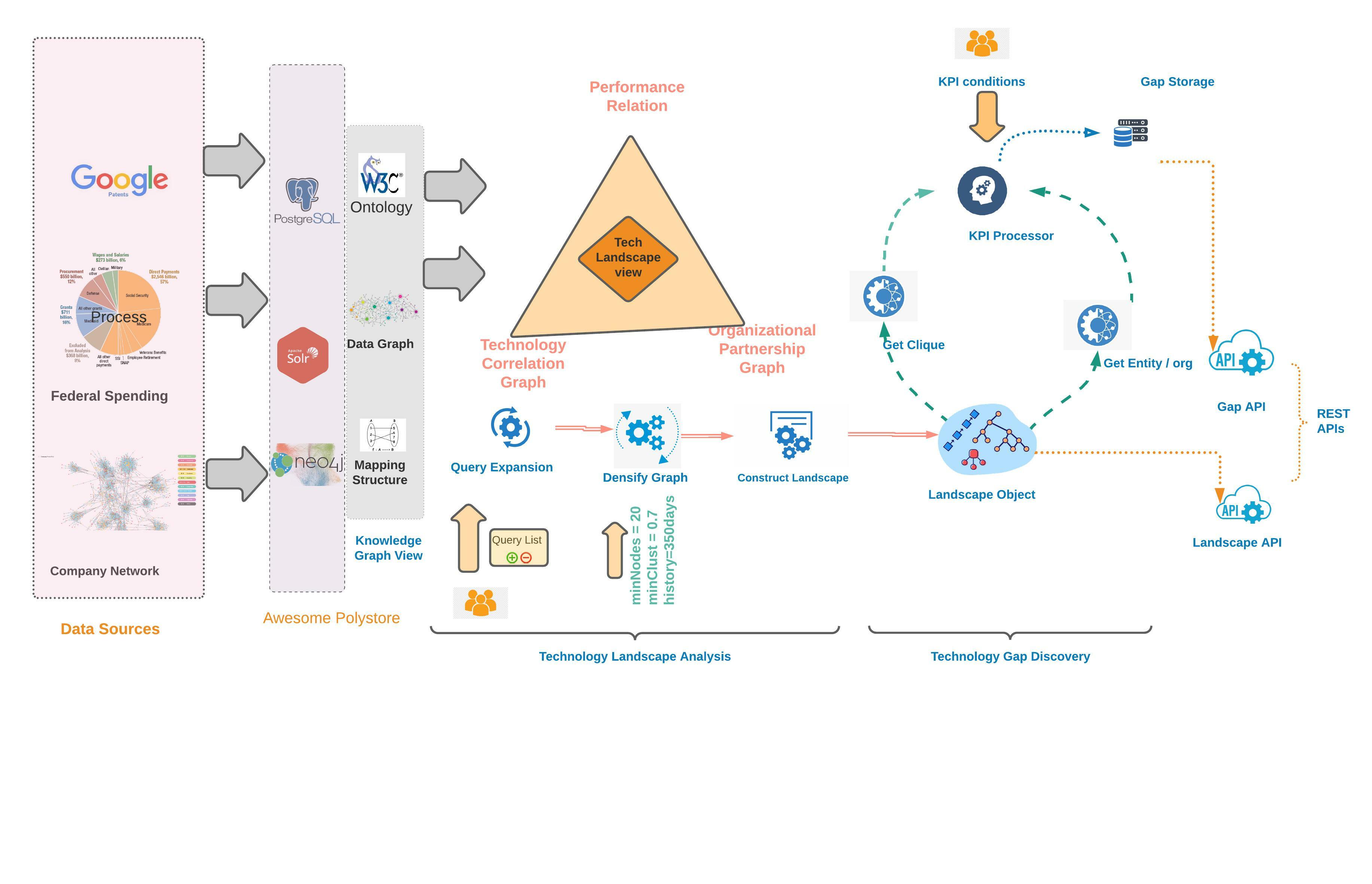}
  \vspace{-2.7cm}
    \caption{The information processing architecture of the System}
    \label{fig:KGstore}
\end{figure*}

\noindent \textbf{Data Graph.} The data graph is constructed from heterogeneous data sources and distributed across all three stores. The knowledge graph construction process is beyond the scope of this document. Table \ref{tab:sources} shows a few of the data sources, their structure, and their placement in the polystore. The knowledge graph is designed as a \textit{materialized polystore view} over these component data sets. Unlike a database view that is defined as a single (potentially complex) query against a set of base tables or views, a polystore view is specified as a query script that specifies the relational, graph and text-index components of the view. For example, consider the patent data source which is partitioned into the patent metadata component residing in a table, and a patent description component, which is stored in Solr. The entities derived from patents include the patents themselves, organizations, individuals, the technology concepts pertaining to the patent; the relationships include co-ownership, IP-transfer, significant co-occurrences between technologies in selected sections of the patent, etc. As the view is materialized, the entities (together with their properties) are stored in separate relational tables, the relationships together with their end entities are stored in the graph database, and additional indices (e.g., one for time and technology terms) are stored in separate index structures. Figure \ref{fig:KGstore} shows a pictorial view of this materialized structure.

\noindent \textbf{Mapping Structures.} The forward mapping structure is a fast lookup structure that captures the relationship between concepts and their instances in the materialized polystore view. This structure takes the form of a key list and three compressed posting lists that contain the IDs of these instances in the three stores. A reverse mapping structure is also maintained for every store, such that a string occurring in a record of that store is mapped to the corresponding concept node in the ontology. For example, strings ``GPU" and "Graphics Processing Unit" map to the same concept node. Note that this mapping is partial, and an algorithmically detected entity occurring in the data may not have a matching concept in the ontology.

The IntSight Knowledge Navigator tool sits on top of the materialized polystore view and performs knowledge operations that are provided through a set of API calls. 

\section{The Technology Gap Discovery Problem}
\label{gapDiscovery}
Given a knowledge graph where technologies are treated as entities, the gap discovery problem involves two major operations (i) technology landscape analysis and (ii) landscape-based gap discovery.

\medskip

\noindent \textbf{Technology Landscape Analysis.} Informally, a technology landscape captures the ``players'' in a set of specified technology areas and their activities. Hence, we define a \textit{technology landscape} as the triple $L=(P, T, C)$ with \\
- \underline{Performance Relation} $P=(Org, Int, Tech, M_1, M_2, \ldots)$ where $M_i$ is a key performance indicator like the number of patent applications of organization $Org$ on the specific technology $Tech$ in time interval $Int$\\
- \underline{Technology Correlation Graph} $T$ whose nodes represent technologies $t \in dom(Tech)$ and edges represent an ontological relationship or a co-occurrence relationship, and\\
- \underline{Organizational Partnership Graph} $C$ whose nodes are organizations $o \in dom(Org)$ and edges indicate if they have a cooperative relationship (e.g., joint patent holders, coauthors) in a technology area $t \in dom(Tech)$.

\medskip
\begin{algorithm}[hbt]
\caption{Technology Landscape Analysis}
\label{alg:landscape}
\begin{algorithmic}[1]
\Procedure{landscape}{$Pos[], Neg[]$}
\State $OntoList \gets QExpand(Pos, Neg, maxD = 8)\;$
\State $ROIs[] \gets densifyingGraph(OntoList, minNodes = 100, minClust = 0.7, history =$ `5 years'$)$
\State $materialize(G)$
\State $lScape \gets ConstructLandscape(ROIs)$
\State
\Return $lScape$
\EndProcedure
\end{algorithmic}
\end{algorithm}
\noindent Algorithm \ref{alg:landscape} presents the steps for the Technology Landscape Analysis.
The analysis starts with a user specifying two lists $Pos$ consisting of terms of interest, and $Neg$ consisting of terms not of interest. Thus, if $Pos = $ [``query processing'', ``accelerator''] and $Neg = $ [FPGA], the use is interested in hardware accelerator technologies (excluding FPGAs) related to query processing tasks. 

\noindent \textit{Query Expansion.} The first step is to use the ontology for query expansion \cite{wu2011study,azad2019new} to collect a larger set of positive terms that cover the desired technology space from the ontology. The nominal algorithm computes the union of the transitive closure (along specific ontological relationships) of all terms in the $Pos$ list and subtracts from it the union of the transitive closure of terms in the $Neg$ list. However, for an ontology over a million nodes, multiple transitive closure operations is very expensive. We use the node-to-path index and the root-to-node labels effectively to perform query expansion efficiently. 

% \begin{algorithm}[t]
% \caption{Technology Landscape Analysis}\label{algo:landscape}
% \begin{algorithmic}[1]
% \Procedure{LandscapeAnalysis}{$Pos,Neg$}
% \State $newTerms \gets qExpand(Pos, Neg)$
% % \While{$r\not=0$}\Comment{We have the answer if r is 0}
% % \State $a\gets b$
% % \State $b\gets r$
% % \State $r\gets a\bmod b$
% % \EndWhile\label{euclidendwhile}
% % \State \textbf{return} $b$\Comment{The gcd is b}
% \EndProcedure
% \end{algorithmic}
% \end{algorithm}

\noindent \textit{Densifying Subgraph Detection.} With the expanded set of ontological terms $T$, we identify nodes $N$ in the data graph $D$ that correspond to these concept terms using the mapping structures. Using each node in $N$, we identify subgraphs (regions of interest) around these nodes that satisfy the following conditions (a) the number of nodes in the subgraph exceeds $minNodes$, (b) the average clustering coefficient of the subgraph exceeds $minClust$, and (c) the period over which there is a monotonic increase in density matches or exceeds the value of the $history$ parameter. The densifying regions, collectively called the ROI graph $G$, represent parts of the original data graph that are seeing significant activity in recent times. To facilitate the computation of densifying subgraphs, we compute a temporal index of communities in the graph computed through truss-based clustering computed in an HPC cluster. The index is updated every time the materialized view is updated at fixed time intervals. Further, each node in our knowledge graph always maintains some basic network properties about itself including its indegree, outdegree and clustering coefficient. 

\noindent \textit{Landscape Construction.} The technology nodes in the regions of interest are connected to Organization names that have, additional information about them in the entity tables of of the knowledge graph. Using the ROI graph $G$, we extract the companies inside $G$ or within the first neighborhood of $G$ (since each technology term is connected to any organization working on it). For each of these companies, the technologies produced by these companies that are included in $G$ and a set of KPI metrics (e.g., number of patents/publications, $\ldots$) associated with these companies. Secondly, we extract the subgraph induced by the technology nodes from $G$ and merge it with subgraphs from the ontology containing these nodes to create the Technology Correlation graph $T$. Similarly, we extract the subgraph induced by the companies in $G$ and its first neighbors to create the Organization Correlation Graph $C$. The technology landscape thus created is returned to the user which the densifying graph $G$ is materialized for later reuse.

\begin{algorithm}[ht]
\caption{Technology Gap Analysis}
\label{alg:competition}
\begin{algorithmic}[1]
\Procedure{Gap}{$lScape.C, C, cond[], me, \theta$}
 \State $s[] \gets \emptyset$
\State $\gamma \gets (G - Build_{EGO}(lScape.C, me))\;$
\State $\alpha \gets GetQClique(G)$
\State $comp \gets lScape.orgs \cap \gamma.orgs$
\State $\alpha_{org} \gets getOrg(\alpha)$
\For{$n \in comp$}
    \If {$n \in \alpha_{org}$}   
       \If {$ORule(n, cond.org)$}{}
       \If {$TRule(n.qCliques, cond.clique)$}
       \If {$KPIDist(n.KPI, me.KPI) > \theta$}
            \State {$s.append(n, n.qCliques)$}
        \EndIf    
        \EndIf
       \EndIf
    \EndIf   
\EndFor
\State
\Return $s$
\EndProcedure
\end{algorithmic}
\end{algorithm}
\noindent \textbf{Gap Discovery.} The gap discovery method presented in Algorithm \ref{alg:competition}, accepts the landscape and the materialized ROI graph from Algorithm \ref{alg:landscape}, a reference organization $me$ against which the gap will be computed, and a set of predicates $cond[]$ explained below. The algorithm starts by computing the ego network of the $me$ organization to identify all partner institutions. All other organizations are considered as competitors, labeled as $comp$. Independently, the ROI graph $G$ is mined for quasi-cliques $\alpha$ to identify technology combinations that have dominant activity in which the members of $comp$ participate. These participating organizations, labeled as $\alpha_{org}$, are filtered based on three sets of conditions: 1) The organizations must satisfy a set of organization specific properties (e.g., amount of investment), 2) The technology areas they are working on must satisfy certain properties ( major activities in the area must not be less than a year old) and 3) the KPI distance between $me$ and the organization should be beyond a threshold $\theta$. The KPI, are computed from the $M_i$ properties of the organization in the Performance Relation $P$ of the technology landscape. The result of the algorithm is the organization's information from $P$, and quasi-clique in $\alpha$ that the organization is associated with. 
\section{The IntSight Knowledge Navigator}
\label{sec:IKN}
\begin{figure}
    \centering
    \includegraphics[width=0.45\textwidth]{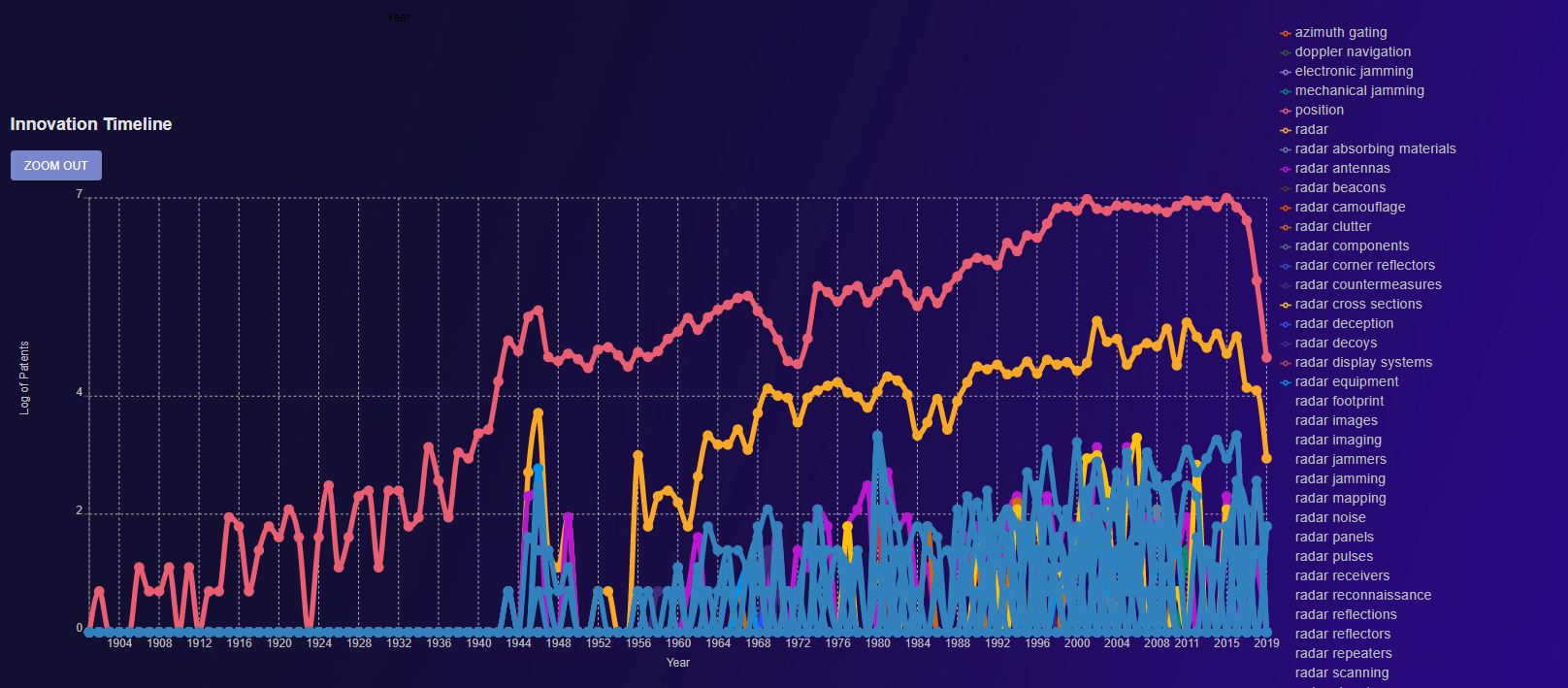}
    \caption{Comparative timeline of innovations in the technology area in the specified time interval.}
    \label{fig:timeline}
\end{figure}

\begin{figure}[t]
    \centering
    \includegraphics[width=0.45\textwidth]{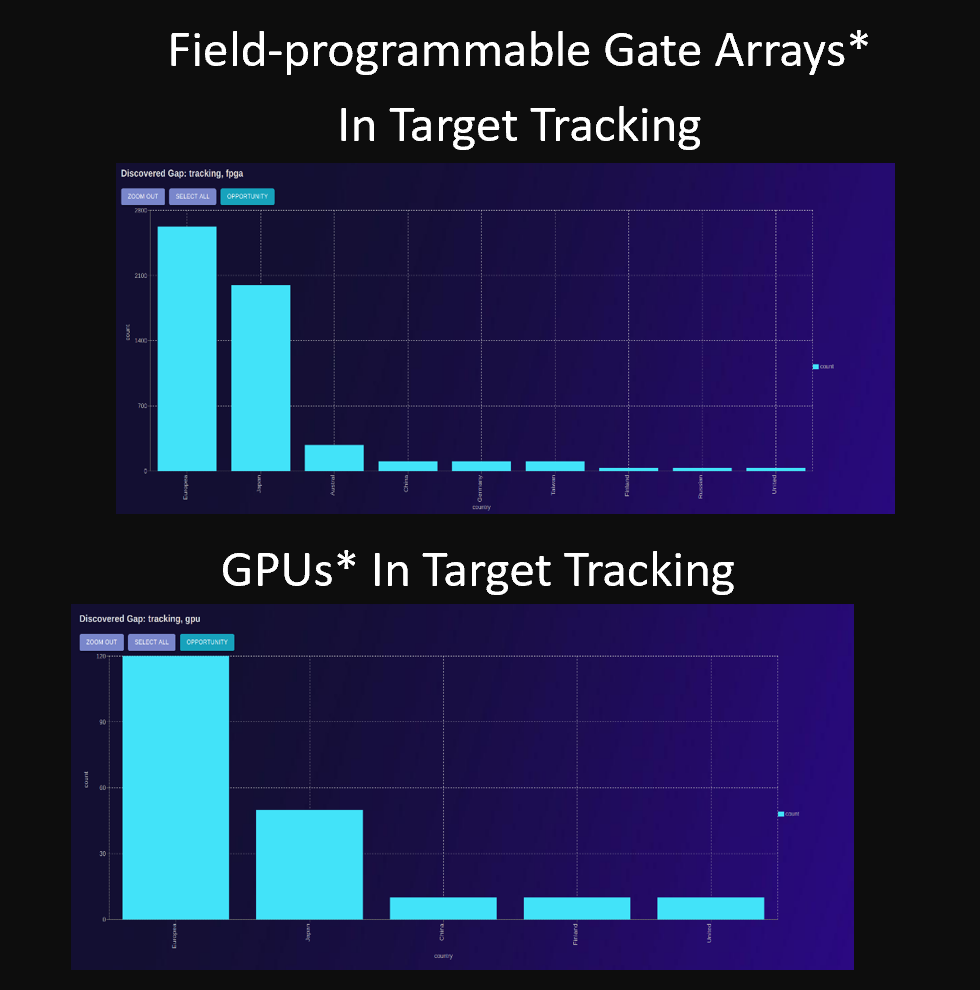}
    \caption{A comparative analysis of gaps for two technologies. The bar charts on the left are the leading organization in the technology area.}
    \label{fig:comparative}
\end{figure}

The IntSight Knowledge Navigator is a web-based exploration and analysis tool that enables users to perform the two analyses through a dashboard-like interface. The tool uses the platform API and the Knowledge Graph API. For example, the gap analysis returns the organization ID and its high-activity technology areas. The API invokes a database join operation with the Performance Relation and the Entity Tables to construct the result objects needed for visualization. 
\begin{figure}[t]
    \centering
    \includegraphics[width=0.45\textwidth]{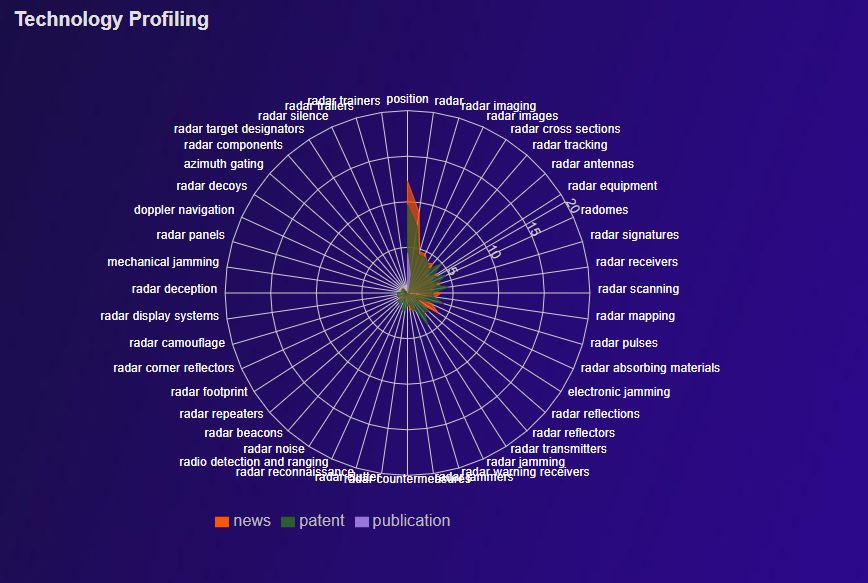}
    \caption{A spider chart showing the result of query expansion and the relative volumes of data from multiple data sources.}
    \label{fig:spider}
\end{figure}
The dashboard shows the results for every step in the process to facilitate human interaction. Figure \ref{fig:spider} shows partial results for the query $Pos=[radar]$. For result elements like the Performance Relation, the API returns the result of a data cube operation and the dashboard displays several different charts for different group combinations that the system finds informative based on factors used in visualization recommender systems \cite{lee2021deconstructing}. For the same query Figure \ref{fig:timeline} shows the timeline of innovations for different technologies related to the query. 
The user can choose to perform comparative gap analysis between organizations and technologies. Figure \ref{fig:comparative} shows the results of a gap capmparison of FPGAs vs. GPUs for a specific organization in the context of target tracking.

\bibliographystyle{IEEEtran}
\bibliography{ref}

\end{document}